# Thermohaline circulation stability: a box model study - Part II: coupled atmosphere-ocean model

Valerio Lucarini *

and

Peter H. Stone †

Department of Earth, Atmospheric and Planetary Sciences, MIT

Cambridge, MA 02139 USA

and

Joint Program on the Science and Policy of Global Change, MIT

Cambridge, MA 02139 USA

September 26, 2004

---

*lucarini@alum.mit.edu; now at Dipartimento di Matematica ed Informatica, University of Camerino, via Madonna delle Carceri 62032, Camerino (MC) Italy

†phstone@mit.edu

**Abstract**

A thorough analysis of the stability of a coupled version of an inter-hemispheric 3-box model of Thermohaline Circulation (THC) is presented. This study follows a similarly structured analysis on an uncoupled version of the same model presented in Part I. The model consists of a northern high latitudes box, a tropical box, and a southern high latitudes box, which respectively can be thought as corresponding to the northern, tropical and southern Atlantic ocean. We study how the strength of THC changes when the system undergoes forcings representing global warming conditions.

Since we are dealing with a coupled model, a direct representation of the radiative forcing is possible, because the main atmospheric physical processes responsible for freshwater and heat fluxes are formulated separately. Each perturbation to the initial equilibrium is characterized by the total radiative forcing realized, by the rate of increase, and by the North-South asymmetry. Although only weakly asymmetric or symmetric radiative forcings are representative of physically reasonable conditions, we consider general asymmetric forcings, in order to get a more complete picture of the mathematical properties of the system. The choice of suitably defined metrics allows us to determine the boundary dividing the set of radiative forcing scenarios that lead the system to equilibria characterized by a THC pattern similar to the present one, from those that drive the system to equilibria where the THC is reversed. We also consider different choices for the atmospheric transport parameterizations and for the ratio between the high latitude to tropical radiative forcing. We generally find that fast forcings are more effective than slow forcings in disrupting the present THC pattern, forcings that are stronger in the northern box are also more effective in destabilizing the system, and that very slow forcings do not destabilize the system whatever their asymmetry, unless the radiative forcings are very asymmetric and the atmospheric transport is a relatively weak function of the meridional temperature gradient. In this latter case we present some relevant hysteresis graphs of the system. The changes in the strength of the THC are primarily forced by changes in the latent heat transport in the hemisphere, because of its sensitivity to temperature that arises from the Clausius-Clapeyron relation.

# 1. Introduction

This analysis follows the study reported in Part I of this paper (Lucarini and Stone 2004), where we have dealt with an uncoupled 3-box oceanic model. Therefore, the reader should refer to the introduction of Part I for a brief presentation of the concept of thermohaline circulation (THC) and for a discussion of the relevant issues regarding the THC's fate in the context of global warming as well in a paleoclimatic perspective.

The model here analyzed is characterized by the presence of explicit coupling between the ocean and the atmosphere. The atmospheric freshwater and heat fluxes are expressed as functions of the oceanic temperatures. We emphasize that the analytic formulation of the freshwater flux includes the Clausius-Clapeyron effect, while the total atmospheric heat flux is given by the sum of three functionally distinct contributions, representing the latent, sensible, and radiative heat fluxes. The more detailed mathematical description of the heat transfer processes allows us to consider a conceptually more precise picture of global warming scenarios where the external radiative forcing acts on the radiative heat flux term alone. Previous studies have shown that the interaction with the atmosphere often has a strong effect on the stability properties of the THC (Nakamura et al. 1994; Scott et al. 1999; Wang et al. 1999b).

In this paper we perform a parametric study of the relevance of both spatial *and* temporal patterns of the radiative forcing and characterize the response of the system by determining the thresholds beyond which we have destabilization of the present mode of the THC and transition to a reversed circulation. We consider an extremely simplified climate model. On one side, this allows a very thorough exploration of the parameters' space. On the other side, this limits the scope of this paper to providing qualitative information which should be considered as conceptual and methodological suggestions for more detailed analysis to be performed using more complex models.

Our paper is organized as follows. In section 2 we provide a brief description and the general mathematical formulation of the dynamics of the three-box model used in this study and we describe the parameterization of the coupling between the atmosphere and the ocean. In section 3 we describe the forcings we apply. In section 4 we analyze some relevant model runs. In section 5 we present a study of some relevant hysteresis graphs of the system. In section 6 we treat the general stability properties of the system. In section 7 we perform a parametric sensitivity study. In section 8 we present our conclusions.

## 2. Brief model description

The three-box model consists of a northern high latitude box (box 1), a tropical box (box 2), and a southern high latitude box (box 3). The volume of the two high latitudes boxes is the same, and is $1/V$ times the volume of the tropical box. We choose $V = 2$, so that box 1, box 2 and box 3 respectively can be thought as describing the portions of an ocean like the Atlantic north of $30°N$, between $30°N$ and $30°S$, and south of $30°S$. The two high latitude boxes are connected by a deep current passage of negligible mass. The boxes are well mixed, and $T_i$ and $S_i$ are respectively the temperature and salinity of the box $i$, while $\tilde{F}_i$ and $\tilde{H}_i$ are respectively the net freshwater and heat atmospheric fluxes into box $i$. The box $i$ is subjected to the oceanic advection of heat and salt from the upstream box through the THC, whose strength is $\tilde{q}$. The atmosphere and the land have a negligible heat capacity and water content compared to the ocean. The structure of the model we consider in this paper is the same of the model analyzed in Part. We suggest the reader to check the description of the model, the notation, the discussion of the tendency equations, and their non-dimensionalization in Part I. We remind that the *transport* quantities without the *tilde* the are suitably non-dimensionalized counterparts of the physical variables.

**a.** *Parameterization of the coupling between the atmosphere and the ocean*

The model we analyze in this work is formulated in a different way with respect to Part I since it incorporates an explicit coupling between the ocean and the atmosphere. The atmospheric fluxes of heat and freshwater are parameterized as functions of the box temperatures. We choose simple but physically plausible functional forms which are based on the large scale processes governing the transfer of heat and freshwater through the atmosphere. We want to capture the dependence of atmospheric transport from the tropics to the high-latitudes on the temperature gradient, considering that baroclinic eddies contribute to most of the meridional transport around $30°N$ (Peixoto and Oort 1992), the dependence of the outgoing long wave radiation on the temperature, and the dependence of the moisture content of the atmosphere on the temperature. The last property differentiates the parameterization we choose from the otherwise closely similar Scott et al. (1999) model's choices.

**b.** *Freshwater fluxes*

The non-dimensionalized net freshwater fluxes $F_i$ are parameterized following Stone and Miller (1980) and Stone and Yao (1990):

$$F_i = \frac{C_i}{\left(\frac{T_2+T_i}{2}\right)^3} \exp\left[-\frac{L_v}{R_v \frac{T_2+T_i}{2}}\right] (T_2 - T_i)^n, \quad i = 1, 3 \tag{1}$$

$$F_2 = -\frac{1}{V}(F_1 + F_3) \tag{2}$$

where $L_v$ is the unit mass latent heat of vaporization of water (taken as constant), $R_v$ is the gas constant, and $C_1$ and $C_3$ are coefficients we have to calibrate. The exponential functions are derived from the Clausius-Clapeyron law, while the value of the exponent $n$ determines how sensitive is the process of baroclinic eddy transport on the meridional temperature gradient. $F_2$ is obtained as in Part I by imposing the conservation of the salinity.

We note that the Clausius-Clapeyron equations had been included in the description of atmosphere-ocean coupling in earlier studies performed on box models, but those dealt with hemispheric and not inter-hemispheric models (Nakamura et al. 1994; Tziperman and Gildor 2002).

**c.** *Heat fluxes*

The surface heat fluxes $\tilde{H}_i$ are decomposed in three components describing physically different phenomena:

$$\tilde{H}_i = \tilde{LH}_i + \tilde{SH}_i + \tilde{RH}_i \qquad (3)$$

where $\tilde{LH}_i$ and $\tilde{SH}_i$ are respectively the convergence of the atmospheric flux of latent and sensible heat in the box $i$, while $\tilde{RH}_i$ describes the radiative balance between incoming solar radiation and outgoing longwave radiation. The convergence of atmospheric transports must globally sum up to zero at any time, since the atmosphere is closed.

**i. Latent heat flux** The oceanic box $i = 1, 3$ receives a fraction $1/\gamma_i$ of the total net moisture transported by the atmosphere from the tropics to high latitudes in the northern ($i = 1$) and southern hemisphere ($i = 3$) respectively; this includes the fraction that directly precipitates over the oceanic boxes $i = 1, 3$ and the fraction that precipitates over land and runs off to the oceanic box $i = 1, 3$.

The fractional catchment area $1/\gamma_i$ can range in our system from $1$ (all of the atmospheric moisture exported from the tropics to the high latitudes ends up respectively in the box $i = 1, 3$) to $1/6$ (the box $i = 1, 3$ receives only the moisture transported from the tropics that precipitates on the ocean surface). The remaining fraction $(1 - 1/\gamma_i)$ of the total atmospheric moisture exported from the tropics returns back to the oceanic box 2 by river runoff or underground flow. This latter fraction does not affect the moisture budget of the oceanic boxes $i = 1, 3$ but does affect their heat budget, since the process of condensation occurs over the high latitude regions $i = 1, 3$, so that latent heat is released to the atmosphere and is immediately transferred to the oceanic box $i = 1, 3$. We then obtain that freshwater and the latent heat fluxes are related as follows:

$$\tilde{LH}_i = \gamma_i \cdot L_v \cdot \tilde{F}_i, \quad i = 1, 3, \tag{4}$$

$$\tilde{LH}_2 = -\tilde{LH}_1 - \tilde{LH}_3, \tag{5}$$

where we have used that the total net balance of the latent heat fluxes is zero. When we apply the non-dimensionalization procedure as in Part I, we obtain that the following relation holds between $LH_i$ and $F_i$:

$$LH_i = \gamma_i \frac{L_v}{c_p \cdot S_0 \cdot \rho_0} F_i, \quad i = 1, 3. \tag{6}$$

We observe that in density units the following relation holds:

$$\left(\frac{LH_i}{F_i}\right)_\rho = \frac{\alpha \cdot \gamma_i \cdot L_v}{\beta \cdot c_p \cdot S_0} \approx 6, \quad i = 1, 3. \tag{7}$$

We underline that this value of the ratio depends on our assumed values for $\gamma_i$, $\alpha$ and $\beta$. If we consider a more realistic equation of state for the density, the relative importance of freshwater and heat fluxes is expected to change in quantitative - but not qualitative - terms with respect to what presented in equation (7).

**ii. Sensible heat flux** We consider that the baroclinic eddies are the main mechanism responsible for the meridional transport of sensible heat, therefore we assume that the sensible heat flux convergence $\tilde{SH}_i$ is, coherently with our picture of the latent heat transport, proportional to the $n^{th}$ power of meridional temperature gradient (Stone and Miller 1980; Stone and Yao 1990). After the suitable non-dimensionalization, we obtain the following expression for $SH_i$:

$$\begin{aligned} SH_i &= D_i \left(T_2 - T_i\right)^n, \quad i = 1, 3 \\ SH_2 &= -\frac{1}{V} \left(SH_1 + SH_2\right), \end{aligned} \tag{8}$$

where $SH_2$ is such that the total sensible heat flux convergence globally sum up to zero.

**iii. Radiative heat flux** The radiative heat flux $\tilde{RH}_i$ is modelled as usual as a newtonian relaxation process (Wang and Stone 1980; Marotzke and Stone 1995; Marotzke 1996). After the correct non-dimensionalization, we obtain the following expression for $RH_i$:

$$RH_i = A_i - B_i T_i = B_i \left(\frac{A_i}{B_i} - T_i\right) = B_i \left(\vartheta_i - T_i\right), \quad i = 1, 2, 3. \tag{9}$$

With respect to the box $i$, $A_i$ describes the net radiative budget if $T_i = 0°C$, $B_i$ is an empirical coefficient, which, if albedo is fixed, as in this model, is a measure of the sensitivity of the thermal emissions to space to surface temperature, including also the water vapor and clouds feedbacks, and $\vartheta_i = A_i/B_i$ is the radiative equilibrium temperature.

**iv. Choice of the constants** In order to obtain for the coupled model here presented an initial equilibrium identical to that of the uncoupled model, which is summarized in the data reported in table 2 of Part I, we need to carefully choose the constants in the atmospheric parameterization. We first set the radiative heat flux parameters by adopting the parameterization proposed by Marotzke (1996). We set for our model $B_1 = B_2 = B_3 = B = 5.1 \cdot 10^{-10} \ s^{-1}$. We can then obtain the the following restoring equation for the global average temperature $T_M$:

$$\dot{T}_M = B(\vartheta_M - T_M); \tag{10}$$

where the parameter $B$ introduces a time scale of $\approx 60$ years for the global radiative processes (Marotzke 1996; Scott et al. 1999). The $\vartheta_i$ are chosen following the parameterization presented in (Marotzke 1996) and are such that the average global radiative temperature $\vartheta_M \equiv (\vartheta_1 + V \cdot \vartheta_2 + \vartheta_3)/(2 + V) = 15°C$, which is also the value of the global average temperature at equilibrium, as in the uncoupled model in Part I. In physical terms, $B$ corresponds to the property of the system that a global radiative forcing of $1 \ Wm^{-2}$ (which results in an effective radiative forcing of $6 \ Wm^{-2}$ in the oceanic surface fraction, since we assume that land and atmosphere have no heat capacity) causes an increase of the average temperature of the system $T_M$ of $\approx 0.6°C$ when equilibrium is re-established; this property can be summarized by introducing a climatic temperature/radiation *elas-* *ticity* parameter $\kappa_M \approx 0.6\ KW^{-1}m^2$. Therefore, considering that it is estimated that in the real Earth system the doubling of $CO_2$ causes an average radiative forcing of $\approx 4\ Wm^{-2}$, we can loosely interpret the parameter $\kappa_M$ as indicating a model *climate sensitivity* of $\approx 2.5\ ^\circ C$.

Substituting in expressions (1) and (2) the $T_i$ of the equilibrium solution of the uncoupled model, we can derive the $C_i$ such that we obtain $F_1 = 13.5 \cdot 10^{-11}\ psu\ s^{-1}$ and $F_3 = 9 \cdot 10^{-11} psu\ s^{-1}$. The latent heat fluxes are then obtained using equation (6), while the coefficients $D_i$ for the sensible heat fluxes in equation (8) are derived by requiring that the total heat flux $H_i$ in (3) of the coupled and of the uncoupled model match at the equilibrium solution. The relative magnitude of the latent and sensible heat fluxes $LH_i$ and $SH_i$ at equilibrium depend on the choice of $\gamma_i$ (Marotzke 1996). Estimates for $\gamma_i$ for the Atlantic range from $1.5$ to $3$ (Marotzke 1996). We set $\gamma_1 = \gamma_3$ in order to keep the geometry of the problem entirely symmetric, and choose $\gamma_1 = \gamma_3 = 2$.

The parameter $n$ determines the efficiency of the atmospheric transports in terms of its sensitivity to the meridional temperature gradient: we consider for $n$ the values $[1, 3, 5]$, which include the domain proposed in (Held 1978) (2 to 5), but also include simple diffusive representation ($n = 1$). We report in table 1 the value of the main model's constants and in table 2 the value of the quantities defining the initial equilibrium state. It is important to note that the initial equilibrium value of the THC strength introduces a natural time scale for the system $q_{eq}^{-1} = t_s \approx 250$ $y$, which is the flushing (or advection) time of the oceanic boxes. In particular we observe that while the sensible and latent heat fluxes into box 1 are roughly the same, in the case of box 3 the sensible heat flux is almost three times as large as the latent heat flux. We note that, in spite of the rather rough procedure of parameters' estimation and of deduction of the sensible and latent heat fluxes at equilibrium, the figures we obtain with our model are broadly in agreement with the climatological estimates (Peixoto and Oort 1992), as can be seen in table 3.

## 3. Feedbacks of the system and radiative forcings applied

The oceanic component of our model contain the well-known oceanic feedbacks, the salinity advection (positive) and heat advection (negative) feedbacks (Scott et al. 1999; Lucarini and Stone 2004). However, the atmospheric feedbacks and how they coupled to the ocean are considerably more complex than in the analysis of Scott et al. (1999). In our case, we have to consider both changes in the meridional temperature gradients $T_i - T_2$ and in the hemispheric temperature averages $(T_i + T_2)/2$, whereas only the former was relevant in the analysis of Scott et al. (1999). In general, we expect that positive changes in the average temperature enhance the amount of moisture that can be retained by the atmosphere, thus increasing the meridional transports of moisture and latent heat. On the other hand, we expect that the reduction of meridional temperature gradients hinders the efficiency of the meridional atmospheric transports.

The complexity of the coupled model makes very cumbersome a detailed schematic description of the feedbacks of the system presented along the lines of (Marotzke 1996; Scott et al. 1999). The main aim of this work is to describe the behavior and the stability properties of a strongly perturbed system. Therefore, instead of analyzing in detail the internal feedbacks of the unforced system, we will go directly to the study of the response of the system to the external forcing in order to capture the most relevant processes. We will then deduce which are the most relevant processes controlling the dynamics of the system by interpreting the sensitivity of the system response to some key parameters.

In our model, we simulate global warming-like radiative forcings by increasing the radiative equilibrium temperatures $\vartheta_i$. Since the dynamics of the model depends on both averages and gradients of temperatures of neighboring boxes, we cannot limit ourselves to considering only changes in high-latitude boxes as in Part I. Therefore, changes in the parameter $\vartheta_2$, which in the first approximation controls $T_2$, need to be considered in order to perform a complete and sensible study. We alter the driving parameters $\vartheta_i$ by using a linear increase:

$$\vartheta_i(t) = \left\{ \begin{array}{ll} \vartheta_i(0) + \vartheta_i^t \cdot t, & 0 \leq t \leq t_0 \\ \vartheta_i(0) + \vartheta_i^t \cdot t_0. & t > t_0 \end{array} \right. \quad i = 1, 2, 3. \quad (11)$$

We make this choice because a linearly increasing radiative forcing approximately corresponds in physical terms to an exponential increase of the concentration of greenhouse gases (Shine et al. 1995; Stocker and Schmittner 1997). The role of $\vartheta_2$ is analyzed by considering three cases of TRopical to high-latitudes Ratio of Forcing $TRRF = \Delta\vartheta_2/\Delta\vartheta_1 = [1.0, 1.5, 2.0]$, where we have used the definition $\Delta\vartheta_i = \vartheta_i^t \cdot t_0$. This allows for the fact that in global warming scenarios the net radiative forcing increase is larger in the tropics than in mid-high latitudes (Ramanathan et al. 1979). Such discrepancy is due to the stronger water-vapor positive feedback in the tropics caused by larger amount of water vapor in the tropical atmosphere.

For a given value of $TRRF$, each forcing scenario can be uniquely identified by the triplet $[t_0, \Delta\vartheta_3/\Delta\vartheta_1, \Delta\vartheta_1]$ or alternatively by the triplet $[t_0, \Delta\vartheta_3/\Delta\vartheta_1, \vartheta_1^t]$. Although only weakly asymmetric or symmetric forcings are representative of physically reasonable conditions, in our study we consider general asymmetric forcings, in order to get a more complete picture of the mathematical properties of the system and so derive a wider view of the qualitative stability properties of the THC system.

We find that in most cases the destabilization of the THC occurs if $\Delta\vartheta_1 \geq \Delta\vartheta_3$. We observe that the first order effect of such a forcing is to weaken the THC, because the following relation holds:

- Increase in $\vartheta_2 \geq$ Increase in $\vartheta_1 \geq$ Increase in $\vartheta_3 \Rightarrow$ Increase in $H_2 \geq$ Increase in $H_1 \geq$ Increase in $H_3 \Rightarrow$ Increase in $T_2 \geq$ Increase in $T_1 \geq$ Increase in $T_3 \Rightarrow q$ decreases.

We underline that in general a larger radiative forcing in the tropical box decreases the stability of the circulation because it causes advection of warmer water from box $2$ to box $1$.

## 4.  Analysis of selected model runs

In figure 1 we show the time evolution of the THC strength $q$ (we have chosen $n = 1$) for two slightly different symmetric radiative forcings lasting $500$ years: the solid line describes the subcritical and the dashed line the supercritical case.

In the subcritical case the forcing is such that the radiative equilibrium temperature increases by 7.5 $°C$ per century in both the high-latitude boxes and by 11.25 $°C$ per century in the tropical box. This corresponds to changes in the radiative forcings of $\approx 12\ Wm^{-2}$ per century and $18\ Wm^{-2}$ per century respectively and so to a globally averaged increase of the radiative forcing of $\approx 15\ Wm^{-2}$ per century.

In the subcritical case, the minimum value of THC strength (which is $\approx 6\ Sv$) is reached at $t \approx 350\ y$. After that, and so still during the increase in the forcing, $q$ oscillates with a period of $\approx 400\ y$. This means that the negative feedbacks overcome the external forcing and stabilize the system. We see that in the case of symmetric *global warming-like* radiative forcing, the short-term response of the system is characterized by an initial decrease of the THC strength, in agreement with the results of most of the models (Tziperman 2000).

We underline that in the case of the uncoupled model, the system closely follows the thermal forcing up to the end of its increase (see figure 7 in Part I). This change in the behavior of the system is essentially due to the presence in the uncoupled model of a much stronger temperature restoring coefficient, which summarizes the effects of all of the various surface heat fluxes.

In order to explore the processes responsible for the internal feedbacks characterizing the stability properties of the system around the initial state, we analyze in detail the diagnostics of the subcritical case, which is more instructive because the negative feedbacks eventually prevail on the external destabilizing perturbations.

In figure 2 we analyze the effect of the radiative forcing on the meridional temperature gradient in parallel to the evolution of the THC. Figure 2b) shows that both the Northern and Southern Hemisphere temperature gradients, after an initial slight increase due to the first-order effect related to the larger direct radiative forcing realized in the tropical regions, present a somewhat unexpected large decrease in response to the climate forcing, with the value of $T_2 - T_1$ reaching at equilibrium a value of $\sim 10°C$. The enhanced warming realized at the high latitudes resemble qualitatively the results of most GCMs in the simulation of global warming (Cubasch et al. 2001). Nevertheless, we need to point out that a major reason for high-latitude warming in these models is the albedo feedback, which we don't have. Moreover, in GCMs' transient experiments, they generally have strong high latitude warming in the NH, but not in the SH, because of the stronger heat uptake in the SH ocean, which is not represented in our model (Cubasch et al. 2001). Figure 2c) shows that in both hemispheres the change of latent heat flux is the process providing the single most relevant contribution to this effect. While the forcing is dominant, the increase in the latent heat flux counteracts the decrease of the oceanic thermal advection caused by the decrease of the THC strength. At the final equilibrium, in both hemispheres the latent heat fluxes are much larger than the corresponding oceanic transports, while at the initial equilibrium the two processes are of comparable size. In the time frame where we observe the largest increase in the THC strength ($t \sim 400y$), we have a very fast decrease of the meridional temperature gradient caused by the great enhancement of the oceanic thermal advection. The sudden flattening of the meridional temperature structure causes a dip in the value of the latent heat fluxes.

Since when a northern sinking equilibrium is realized the value of the THC is monotonically increasing with the value of the freshwater flux in the upwelling box (Rahmstorf 1996; Scott et al. 1999; Lucarini and Stone 2004), the increase of the value of $F_3$ (which is linear with $LH_3$) explains why we obtain at equilibrium a larger value for the THC strength. We wish to point out that the presence of an increase of the THC on a longer term agrees with some more complex models's results (Wang et al. 1999a; Wiebe and Weaver 1999).

In figure 3 we analyze the oceanic and atmospheric buoyancy forcings to the THC in parallel to the evolution of the THC. Such forcings are obtained as difference between the buoyancy forcings in box $1$ and $3$. In figure 3b) we present the evolution of the oceanic advective contributions to the tendency of the THC strength in units of the absolute value of the equilibrium thermal advective contribution. We observe that when the THC is forced to decrease, both the haline and thermal contributions counteract such change. While the negative advective thermal feedback is common in box models, the salinity advective feedback is usually positive. The peculiar behavior we observe for the haline part is due to the fact that during the radiative forcing the moisture balance of the equatorial region is so negative that the northward salinity advection does not decrease. When the neg-ative feedbacks of the system overcome the external forcing and cause a sudden large increase in the THC strength, the haline feedback and the thermal feedback get their usual signs, as can be observed in the spike-like structures for $t \sim 400y$. Moreover, after the end of the forcing, the thermal and haline contributions tend to be out of phase. In figure 3c) we present the evolution of the atmospheric contributions to the forcings to the THC strength in the same units as in panel b). It is clear that the latent heat fluxes play the most relevant role in the destabilization of the system for $t \leq 400y$, since they contribute to all of the variation from the initial value of the atmospheric thermal forcing. The contribution related to the change of the freshwater fluxes is smaller than that provided by the changes of the latent heat fluxes by a factor of $\sim 6$ (see equation (7)). Therefore the largest contribution to the weakening of the THC is thermic and driven by the latent heat.

Two studies with coupled atmosphere-ocean GCMs have obtained contrasting results about the relative relevance of heat *vs.* freshwater as destabilizing mechanisms; in Mikolajewicz and Voss (2000) it is shown that the heat flux change is the most important destabilizing mechanism and that this change is dominated by the latent heat flux, just as in our model. By contrast, in Dixon et al. (1999) it is concluded that changes in the moisture flux were the main destabilizing agent. However, calculations of the contributions of changes in both the heat and moisture fluxes to the change in the density flux in the model used in Dixon et al. (1999) show that it is in fact dominated by the heat flux changes (Huang et al. 2003). Indeed this is true of all the seven coupled GCMs analyzed in the context of the Climate Intercomparison Project (CMIP) (Huang et al. 2003). This apparent con-tradiction may be explained by the results presented in (Kamenkovich et al. 2003), where it is found that, even though the decrease in the thermohaline circulation may be initiated by changes in the moisture flux, changes in the heat flux induced by atmospheric feedbacks nevertheless contribute strongly to the decrease.

We find that, in agreement with other studies considering either relatively sophisticated (Stocker and Schmittner 1997; Schmittner and Stocker 1999) or extremely simplified models (Tziperman and Gildor 2002), that the evolution of latent heat and freshwater meridional fluxes under global warming scenarios depends on two competing effects: the reduced efficiency in the atmospheric transport due to decreased meridional temperature gradient and increased capability of the atmosphere to retain moisture for higher average temperatures. The second effect is thought to dominate for larger climate changes, as shown by paleoclimatic data and simulations of the last few hundred thousand years (Charles et al. 1994; Manabe and Stouffer 1994; Krinner and Genthon 1998; Kitoh et al. 2001), and it is reasonable to expect that similarly the Clausius-Clapeyron effect dominates the changes in the latent heat and freshwater fluxes in global warming scenarios (Tziperman and Gildor 2002).

We briefly analyze how sensitive is the dynamics of the system with respect to the choice of the eddy transport power law by imposing the same forcing as the subcritical case previously analyzed to the system obtained when the value of $n$ is set to $3$ and $5$. In figure 4a) we show the realized evolution of $q$ in the three cases of $n = 1, 3, 5$. We see that the extent of the decline of the THC strength and its length are negatively correlated with the value of the exponent for the eddy transport power law. We have previously analyzed the prominent role of the changes in the latent heat fluxes in determining the dynamics of the system in the $n = 1$ case, so that we limit the comparison to this process.

Anyway, we wish to emphasize that also in the $n = 3$ and $n = 5$ cases we observe the decrease in the meridional temperature gradient in spite of a stronger tropical forcing (not shown). Analogously, we note that in all cases analyzed the final value of the THC is larger than the initial value, and that this positively correlates with the positive difference between the final and initial values of $LH_2$ (not shown). Therefore, such features seem robust.

In figure 4b) we show that the behavior of the latent heat contribution changes with $n$ coherently with the previously presented picture. We have that the timing of the *minima* of the latent heat contributions coincide in all cases with the timing of the THC strength *minima*, and the same applies for the *maxima*. Moreover, comparing the three cases, the *minima* of the latent heat contributions are larger in absolute value when the corresponding *minima* of the THC strength are deeper. Our results suggest that a more temperature-gradient sensitive atmospheric transport is more effective in limiting the destabilizing processes, of which the changes in the latent heat fluxes are the most prominent.

## 5. Hysteresis

Choosing quasi-static perturbations to the radiative temperatures can lead to the reversal of the THC only if we select $n = 1$, the main reason being that if the atmospheric transport is more sensitive to the meridional temperature gradient the system is always able to counteract a slowly increasing destabilizing radiative forcing. This qualitative difference between the behavior of the various versions of the model is independent of the selected value of the parameter $TRRF$. With the choice of $n = 1$ and $TRRF = 1.5$, we have that quasi-static perturbations are destabilizing only if $\Delta\vartheta_3/\Delta\vartheta_1 \leq 0.5$. We present in figures 5 the hysteresis graphs relative to $\Delta\vartheta_3/\Delta\vartheta_1 = [0, 0.1, 0.2, 0.3, 0.4, 0.5]$. In a) we have $\Delta\vartheta_1$ as abscissa and $q$ in units of the initial equilibrium value as ordinate, so that the initial state is the point $(0, 1)$. The abscissae of the bifurcation points on the right hand-side of the graph increase with increasing value of $\Delta\vartheta_3/\Delta\vartheta_1$ from $\approx 12°C$ to $\approx 25°C$, while the ordinates are close to $1$, thus implying that when the bifurcation occurs the value of the THC is only slightly different from the initial equilibrium value. The bistable region is remarkably large in all cases, the total extent increasing with increasing value of $\Delta\vartheta_3/\Delta\vartheta_1$, because the abscissae of the bifurcation points in the left hand-side of the graph circuits are in all cases below $-30°C$ and so well within the unphysical region of parameter space. This means that once the circulation has reversed, the newly established pattern is extremely stable and can hardly be changed again. In b) we present the corresponding hysteresis graphs where the abscissa in this case is the value of the realized freshwater flux into box 1 when the radiatively forced system has reached a newly established equilibrium. The initial equilibrium is the point $(1, 1)$: we can see that there is a monotonic 1 to 1 mapping between the two figures in panels 5a) and 5b), apart from a very limited region around the bifurcation point in the right hand side for low values of $\Delta\vartheta_3/\Delta\vartheta_1$. This implies that in our model at equilibrium the changes of the average surface temperature and the changes in the hydrological cycle are positively correlated, as one would expect. We observe that in terms of freshwater flux the bistable region is somewhat smaller than in the uncoupled case (see figure 5 in Part I), thus suggesting a *caveat* in the interpretation of uncoupled models' results.

## 6. Critical Perturbations

In this section we present the set of critical forcings, which divide the forcings that disrupt the present pattern of the THC from those that drive the system to a northern sinking state qualitatively similar to the initial unperturbed state. We extend the analysis of the hysteresis of the system by considering how the temporal pattern of the perturbation influences the ability of the system to counteract destabilizing forcings. In figure 6 we consider the *model+forcing* case characterized by $n = 1$ and $TRRF = 1.5$. In panel a) of figure 6 we present the manifold of those critical forcings using the coordinate system $[t_0, \Delta\vartheta_3/\Delta\vartheta_1, \Delta\vartheta_1]$, while in panel b) of figure 6 we adopt the coordinate system $[t_0, \Delta\vartheta_3/\Delta\vartheta_1, \vartheta_1^t]$. We present the corresponding results for $n = 3$ and $n = 5$ in figures 7 and 8, respectively. We adopt in all cases a logarithmic scale since, acknowledging the limitations of our model, we are mainly interested in capturing the qualitative properties of the response of the system. There is a general agreement between the response of the coupled and of the uncoupled model (Lucarini and Stone 2004) to destabilizing perturbations. All the figures 6-8 suggest that the more symmetric and the slower the forcing, the less likely is the destabilization of the THC:

- for a given $t_0$, the lower is the value of the ratio $\Delta\vartheta_3/\Delta\vartheta_1$, the lower is the total change $\Delta\vartheta_1$ needed to obtain the reversal of the THC;

- for a given value of the ratio $\Delta\vartheta_3/\Delta\vartheta_1$, more rapidly increasing perturbations (larger $\vartheta_1^t$) are more effective in disrupting the circulation.

We need to point out a very relevant feature which shows how critical is the choice of the parameter $n$. In the $n = 1$ case we have that, consistently with the study of the hysteresis runs presented in figure 5, very slow perturbations can cause the collapse of the THC for low values of $\Delta\vartheta_3/\Delta\vartheta_1$. On the contrary, we have that in the $n = 3$ and $n = 5$ cases there is even for very low values of $\Delta\vartheta_3/\Delta\vartheta_1$ a threshold in the rate of increase of the forcing below which the reversed THC does not occur, independently of the total radiative forcing realized. This can be more clearly understood by observing that in panels b) of figures 7-8 for each value of $\Delta\vartheta_3/\Delta\vartheta_1$, the critical value of the rate of increase of the forcing $\vartheta_1^t$ is independent of the temporal extension of the forcing $t_0$ for large values of $t_0 \geq t_s$. In the $n = 3$ and $n = 5$ cases the system cannot make transitions to a southern sinking equilibrium for quasi-static perturbations, since they would require indefinitely large perturbations.

Previous studies focusing on more complex models obtain a similar dependence of thresholds on the rate of increase of the forcings (Stocker and Schmittner 1997; Schmittner and Stocker 1999), while in other studies where the full collapse of the THC is not obtained, it is nevertheless observed that the higher the rate of increase of the forcing, the larger the decrease of THC realized (Stouffer and Manabe 1999). Other studies on coupled models also show how the spatial pattern of freshwater forcing due to global warming is extremely relevant especially in the short time scales: only forcings occurring mainly in the Northern Atlantic are efficient in destabilizing the THC (Rahmstorf 1996; Rahmstorf and Ganopolski 1999; Manabe and Stouffer 2000; Ganopolski et al. 2001).

## 7. Sensitivity Study

From the analysis of the figures 1, 3, 2, and 4 we have hinted that the difference between the latent heat fluxes into the two high-latitude boxes dominates the dynamics of the forced system and determines its stability. Given the properties and the functional form of $LH_1$ and $LH_3$ we expect that:

1. the system is less stable against radiative forcings if the tropical to high latitudes radiation forcing ratio is larger

2. in the case when the radiative forcing is larger in the northern high-latitude box, the system is more stable if the atmospheric transport feedback is stronger; this effect is likely to be notable *only if* the perturbations have times scales larger than the flushing time of the oceanic boxes, as confirmed in the extreme case of quasi-static perturbations;

To obtain a more quantitative measure of which processes are important, we analyze the sensitivity of the vertical coordinate of the the manifold of the critical perturbations shown in figure 6 to changes in the two key parameters $TRRF$ and $n$. We define $Z_C(t_0, \Delta\vartheta_3, TRRF, n)$ as the function giving the critical value of the change in the radiative temperature of box 1 thoroughly described in the previous section.

Figure 9 shows the value of the 2-dimensional field of the finite differences $\log_{10}[Z_C(t_0, \Delta\vartheta_3, TRRF = 1.0, n = 3)] - log_{10}[Z_C(t_0, \Delta\vartheta_3, TRRF = 2.0, n = 3)]$. We observe that the field does not remarkably depend on the perturbation length $t_0$; this suggests that processes controlling the dynamics of the systems acting on very different time scales of the system are similarly affected by changes in $TRRF$. This means that the increase in the ratio between the tropical and the northern high latitudes radiative forcing changes the response of the system and favors the collapse of the THC evenly and independently of the temporal scale of the forcing itself, and is particularly effective if we use quasi-symmetric forcings. Therefore the change of the parameter $TRRF = \Delta\vartheta_2/\Delta\vartheta_1$ does not have preferential effect on any of the feedbacks.

Figure 10 presents the value of the 2-dimensional field of the finite differences $\log_{10}[Z_C(t_0, \Delta\vartheta_3, TRRF = 1.5, n = 5)] - log_{10}[Z_C(t_0, \Delta\vartheta_3, TRRF = 1.5, n = 3)]$. The most striking feature of this field is that there is a very strong gradient only along the $t_0$ direction and for $t_0 \approx t_s$. The field is small and positive for forcings having temporal scale shorter than the characteristic oceanic time scale of the system, while it becomes very large and positive for forcings having long tem-poral scales. The positive value means that a more sensitive atmospheric transport (higher values of $n$) tends to stabilize the system if $\Delta\vartheta_3/\Delta\vartheta_1 \leq 1$. With a more temperature-gradient sensitive atmospheric transport smaller changes in temperature gradients between the boxes are needed (see figure 2) to change *all* the atmospheric fluxes; therefore the system is able to dynamically arrange very effectively the atmospheric fluxes, so that they can counteract more efficiently the external forcings and keep the system as close as possible to the initial state. If the forcings are very fast, the enhancement of the atmospheric stabilizing mechanism is not very effective. On the contrary, for slow forcings involving time scales comparable to or larger than those of the system, the enhanced strength of the negative feedback obtained with increasing efficiency of the atmospheric transport can play a very significant stabilizing role in the dynamics of the system. We have chosen the $n = 5$ and $n = 3$ difference field, because it gives a clearer picture of the phenomenon we wish to emphasize. Anyway, the previous observations provide an explanation of why the case $n = 1$ is more unstable especially for slow forcings.

These conclusions seems to be in contrast with the result that in several uncoupled models shorter relaxation times for box temperatures - which, as shown by Marotzke (1996), correspond to more sensitive atmospheric heat transports - imply less stability for the system (Tziperman et al. 1994; Nakamura et al. 1994; Scott et al. 1999; Rahmstorf 2000). Actually, the contrast is only apparent, the point being that in an uncoupled model there is no adjustment of moisture fluxes due to temperature changes: therefore decreasing the relaxation time make the system *less flexible* and adaptable to forcings. Moreover, our results seem to dis-agree with the conclusions drawn in the coupled model presented by Scott et al. (1999). We think that the disagreement is mainly due to the fact that in our study we are dealing with a different kind of perturbation, descriptive of global warming, and that these perturbations have a direct and strong influence on the most relevant atmospheric fluxes, because of their highly nonlinear dependence on average temperature changes. This property was not present in the parameterizations used by Scott et al. (1999).

By contrast with the above described result, we see in figure 10 that for $\Delta\vartheta_3/\Delta\vartheta_1 > 1$ and $t_0 \leq 80\ y$, the difference field changes sign. This implies that for extremely fast radiative forcings which are larger in box 3 a system with larger $n$ is more easily destabilized. This occurs because for such forcings the northern meridional temperature gradient is initially forced to increase more than the southern, so that the northward atmospheric fluxes are greatly enhanced. Such process tends to destabilize the circulation and it is stronger for larger values of $n$. Anyway, as observed in Part I, since we have considered the approximation of well-mixed boxes and excluded from the model the equatorial deep box used by other authors Rahmstorf (1996), our model should not be considered very reliable for phenomena taking place in time scales $\ll t_s$.

## 8. Conclusions

In this paper we have analyzed the stability of the THC as described by a set of coupled models differing in the ratio between the radiative forcing realized at the tropics and at high latitudes and in the parameterization of the atmospheric transports.

In a coupled model a natural representation of the radiative forcing is possible, since the main atmospheric physical processes responsible for freshwater and heat fluxes are formulated separately. Although only weakly asymmetric or symmetric radiative forcings are representative of physically reasonable conditions, we have considered general asymmetric forcings, in order to get a more complete picture of the mathematical properties of the system.

We have analyzed five combinations of the system *model+radiative forcing*, considering different combinations of the atmospheric transport parameterization and of the ratio between the high to low latitudes radiative forcing.

When the system is radiatively forced, initially the latent heat fluxes and the freshwater fluxes are strongly enhanced, thanks to the increase in the saturation pressure of the water vapor due to the warming, and the fluxes into the northern high latitude box increase. These are the main causes for the initial reduction of the THC strength. The strong increase of heat flux into box 1 eventually reduces the efficacy of the atmospheric transport and so causes a great reduction of the freshwater flux and of the latent heat (and sensible heat) flux into box 1, which induces an increase in the THC. When the radiative forcings do not drive the system to a collapse, the enhancement of meridional heat fluxes cause large reductions in the meridional temperature gradients, even when larger radiative forcings are prescribed at the tropics. We underline that such a behavior has been obtained even though our model does not include the ice-albedo feedback. This result seems relevant in the interpretation of warm paleoclimates, such as during Eocene.

The variations of latent heat fluxes and of freshwater fluxes into the two high-latitude boxes dominate the dynamics of the forced coupled model. This is mainly due to the very strong nonlinear dependence of water vapor saturation pressure on temperature. Therefore the inclusion of the Clausius-Clapeyron relation in the fluxes' parameterization seems to play a key role in all the results obtained in our study. In our system the major role is played by the latent heat because in density terms it is stronger than the freshwater flux by a factor of $\approx 6$. A qualitatively similar dominance is found in the CMIP models (Huang et al. 2003) and by Mikolajewicz and Voss (2000), while other studies give the opposite result (Stocker and Schmittner 1997; Rahmstorf and Ganopolski 1999; Schmittner and Stocker 1999; Manabe and Stouffer 1999b; Ganopolski et al. 2001).

The hysteresis graphs of the system show formally that that quasi-static perturbations cannot disrupt the northern sinking pattern of the circulation unless we are considering a relatively inefficient atmospheric transport and radiative forcing with large, unrealistic North-South asymmetry. Moreover, in this case the initial equilibrium point is in the bistable region.

We obtain, with a parametric study involving the total forcing realized, its rate of increase, and its North-South asymmetry, the manifold of the critical forcings dividing the forcings driving the system to a southern sinking equilibrium from those that do not qualitatively change the pattern of the THC. We generally find that fast forcings are more effective than slow forcings in disrupting the present THC patterns, forcings that are stronger on the northern box are also more effec-tive in destabilizing the system, and that very slow forcings do not destabilize the system whatever their asymmetry, unless the atmospheric transport is only weakly dependent on the meridional temperature gradient.

We also compute the sensitivity of the results obtained with respect to the tropical-to-high latitude radiative forcing and with respect to the efficiency of the atmospheric transport. A higher forcing in the tropics destabilizes the system evenly at every time scale and greatly favors destabilization for quasi-symmetric forcings. Increasing the efficiency of the atmospheric transports makes the system in general more stable against destabilizing radiative forcings because it allows a very effective control of all the buoyancy fluxes and provides an enhancement of the atmospheric transport's negative feedback. This effect is especially relevant for forcings' time scales comparable or larger than the system's characteristic time scale, while it is not notable for fast forcings, which bypass all the feedbacks of the system.

The adoption of a linear diffusive representation of the atmospheric transport greatly affects the main qualitative aspects of the stability properties of the system. Therefore, one should take great care in adopting such - questionable (Stone and Miller 1980; Stone and Yao 1990) - parameterization in studies where large climatic shift are investigated.

Comparing the results obtained in this study with those presented in Part I, we conclude that the introduction of an explicit atmosphere-ocean coupling increases the stability of the system so that obtaining the collapse of the THC requires very extreme forcings. We conclude that the coupling introduces a negative net feed-back, which is stronger with more efficient atmospheric transports. We can guess that similar effects can be observed with more complex models, thus introducing a caveat in the interpretation of data coming from uncoupled oceanic model.

The relevance of the temporal scale of the forcing in determining the response of our system to perturbations affecting the stability of the THC confirms the findings of Tziperman and Gildor (2002) for a hemispheric coupled box model, of Stocker and Schmittner (1997) and Schmittner and Stocker (1999) in the context of EMICSs, and of (Manabe and Stouffer 1999a), Manabe and Stouffer (1999b), Manabe and Stouffer (2000), and Stouffer and Manabe (1999) in the context of GCMs. The coherence within the whole hierarchical ladder of models gives robustness to this result.

We conclude that when analyzing the behavior of the THC in global warming scenarios with more complex coupled models, the spatial pattern of the radiative forcing should be carefully taken into account, because the THC is a highly nonlinear, nonsymmetric system, and the effect of changing the rate of increase of the radiative forcing should be explored in great detail. This would provide a bridge between the instantaneous and quasi-static changes. The exploration of the THC dynamics in global warming scenarios requires such an approach, since the system encompasses very different time-scales, which can be explored only if the timing of the forcing is varied. Moreover, the sensitivity of our system's response to the parameter $n$ suggests that, in order to capture the behavior of the THC in the context of relevant changes of the climate, it is critical for coupled models to capture precisely the processes responsible for fuelling the the atmospheric transport.

This work allows many improvements, some of which are along the lines of those proposed in Part I. In particular, asymmetries in the oceanic fractional areas would induce asymmetries in the values of $B_i$, thus causing the presence of different restoring times for the various boxes, while asymmetries in the freshwater catchment areas ($\gamma_1 \neq \gamma_3$) would make the relative importance of the latent heat fluxes and of the freshwater fluxes (when expressed in common density units) different in the two boxes $i = 1, 3$.

The presence of the albedo feedback would also enhance any asymmetry between the two high latitudes boxes: it could be included in the model by parameterizing the radiative terms $A_i$ as increasing functions of the temperatures $T_i$, along the lines of Stocker and Schmittner (1997), Schmittner and Stocker (1999), and Tziperman and Gildor (2002), considering the temperatures as proxies for the fraction of the surface covered by ice. We expect that the inclusion of an ice-albedo feedback would decrease the stability of the THC.

The model's reliability would also benefit from the inclusion of a more appropriate nonlinear equation of state for the seawater. A likely effect would be increasing the relative relevance of the freshwater flux changes as driving mechanism of the weakening of the THC.

## Acknowledgments

The authors are grateful to J. Scott for interesting discussions and useful suggestions. One author (V.L.) wishes to thank R. Stouffer for having proposed improve-ments to an earlier version of the manuscript, and T. Stocker and E. Tziperman for having suggested a number of relevant references. This research was funded in part by the US Department of Energy's (DOE) Climate Change Prediction Program and in part by the MIT Joint Program on the Science and Policy of Global Change (JPSPGC). Financial support does not constitute an endorsement by DOE or JPSPGC of the views expressed in this article.

# List of Tables

# List of Figures

| Quantity | Symbol | Value |
|---|---|---|
| Mass of Box $i = 1, 3$ | M | $1.08 \cdot 10^{20}\ kg$ |
| Box 2/Box $i = 1, 3$ mass ratio | V | 2 |
| Average density | $\rho_0$ | $1000\ Kg\ m^{-3}$ |
| Average salinity | $S_0$ | $35\ psu$ |
| Oceanic fractional area | $\epsilon$ | 1/6 |
| Box $i = 1, 3$ fractional water catchment area | $1/\gamma_i$ | 1/2 |
| Specific heat per unit mass of water | $c_p$ | $4 \cdot 10^3\ J\ °C^{-1} Kg^{-1}$ |
| Latent heat per unit mass of water | $L_v$ | $2.5 \cdot 10^6\ J Kg^{-1}$ |
| Gas constant | $R_v$ | $461\ J\ °C^{-1} Kg^{-1}$ |
| Thermal expansion coefficient | $\alpha$ | $1.5 \cdot 10^{-4}\ °C^{-1}$ |
| Haline expansion coefficient | $\beta$ | $8 \cdot 10^{-4}\ psu^{-1}$ |
| Hydraulic constant | k | $1.5 \cdot 10^{-6}\ s^{-1}$ |
| Radiative Equilibrium Temperature - Box 1 | $\vartheta_1$ | $-22.9\ °C$ |
| Radiative Equilibrium Temperature - Box 2 | $\vartheta_2$ | $52.9\ °C$ |
| Radiative Equilibrium Temperature - Box 3 | $\vartheta_3$ | $-22.9\ °C$ |
| Global climatic temperature/radiation elasticity [b] | $\kappa_M$ | $0.6\ °C\ W^{-1} m^2$ |

Table 1: Value of the main model parameters

---

[b]Coupled model - Value relative to the whole planetary surface; corresponds to a *Global climate sensitivity* per $CO_2$ doubling of $\approx 2.5\ °C$

| Variable | Box 1 | Box 2 | Box 3 |
|---|---|---|---|
| Temperature | 2.9 °$C$ | 28.4 °$C$ | 0.3 °$C$ |
| Salinity | 34.7 psu | 35.6 psu | 34.1 psu |
| Atmospheric Freshwater Flux | 0.41 Sv | -0.68 Sv | 0.27 Sv |
| Total Surface Heat Flux | -1.58 PW | 1.74 PW | -0.16 PW |
| -Latent Heat Flux | 1.84 PW | -3.06 PW | 1.22 PW |
| -Sensible Heat Flux | 2.14 PW | -5.75 PW | 3.61 PW |
| -Radiative Heat Flux | -5.57 PW | 10.57 PW | -5.00 PW |
| Oceanic Heat Flux | 1.58 PW | -1.74 PW | 0.16 PW |
| THC strength | 15.5 Sv | 15.5 Sv | 15.5 Sv |

Table 2: Value of the fundamental variables of the system at the initial equilibrium state

| Variable | Box 1 | Box 2 | Box 3 |
|---|---|---|---|
| Latent Heat Flux (Model) | 1.84 PW | -3.06 PW | 1.22 PW |
| Latent Heat Flux (P&O92) | $\sim$ 1.3 PW | $\sim$ -2.6 PW | $\sim$ 1.3 PW |
| Sensible Heat Flux (Model) | 2.14 PW | -5.75 PW | 3.61 PW |
| Sensible Heat Flux (P&O92) | $\sim$ 1.5 PW | $\sim$ -4 PW | $\sim$ 2.5 PW |

Table 3: Comparison between the values of sensible and latent heat fluxes of the model at the initial equilibrium and the corresponding climatological estimates given by Peixoto and Oort (1992) for the net fluxes at 30°N and 30°S (in the table P&O92).

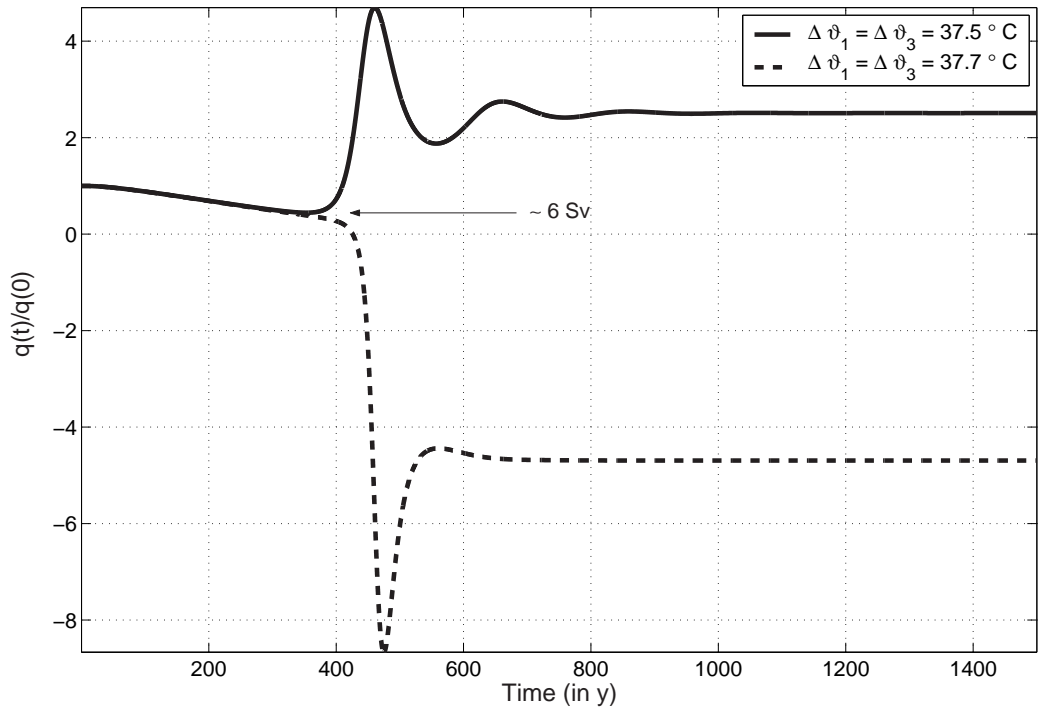

Figure 1: Evolution of the THC strength under a super- and sub-critical radiative flux forcing. In both cases $\Delta\vartheta_2 = 1.5\Delta\vartheta_1 = 1.5\Delta\vartheta_3$.

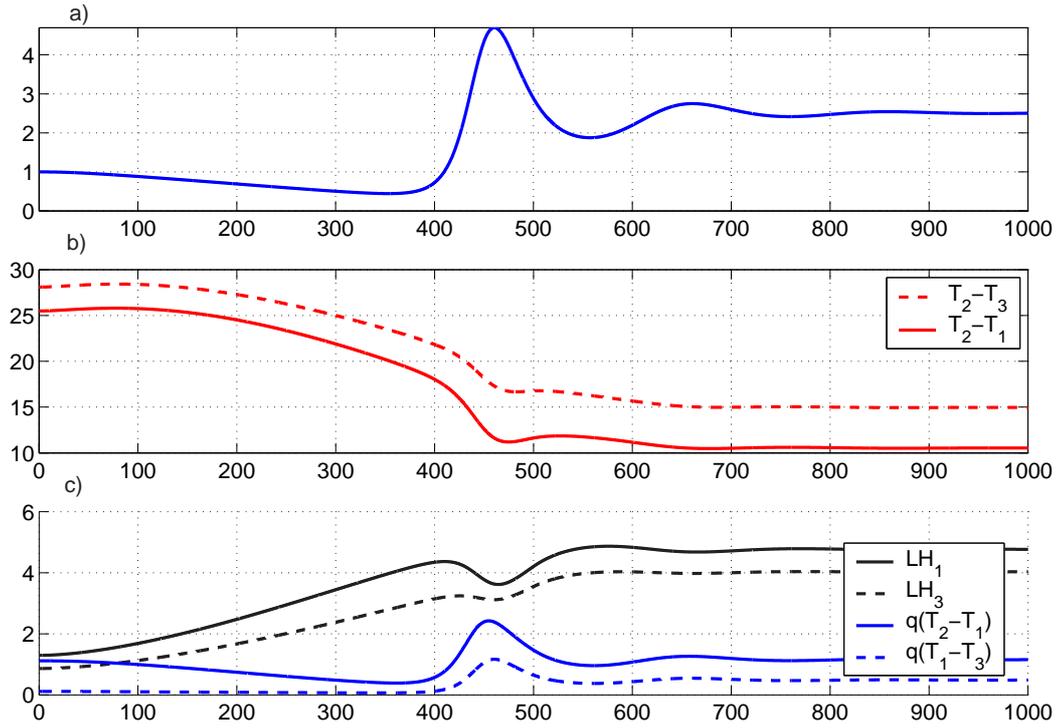

Figure 2: a) Evolution of the THC strength in units of $q(0)$ for the subcritical case presented in figure 1. b) Evolution of the meridional temperature gradient $T_2 - T_1$ and $T_2 - T_3$ in units of $°C$. c) Evolution of the latent heat fluxes and of the oceanic advection thermal forcings $q(T_2 - T_1)$ (into box 1) and $q(T_1 - T_3)$ (into box 3). Units of $|\alpha q(0)(T_2(0) - 2T_1(0) + T_3(0))|$. Abscissae: elapsed time expressed in $y$.

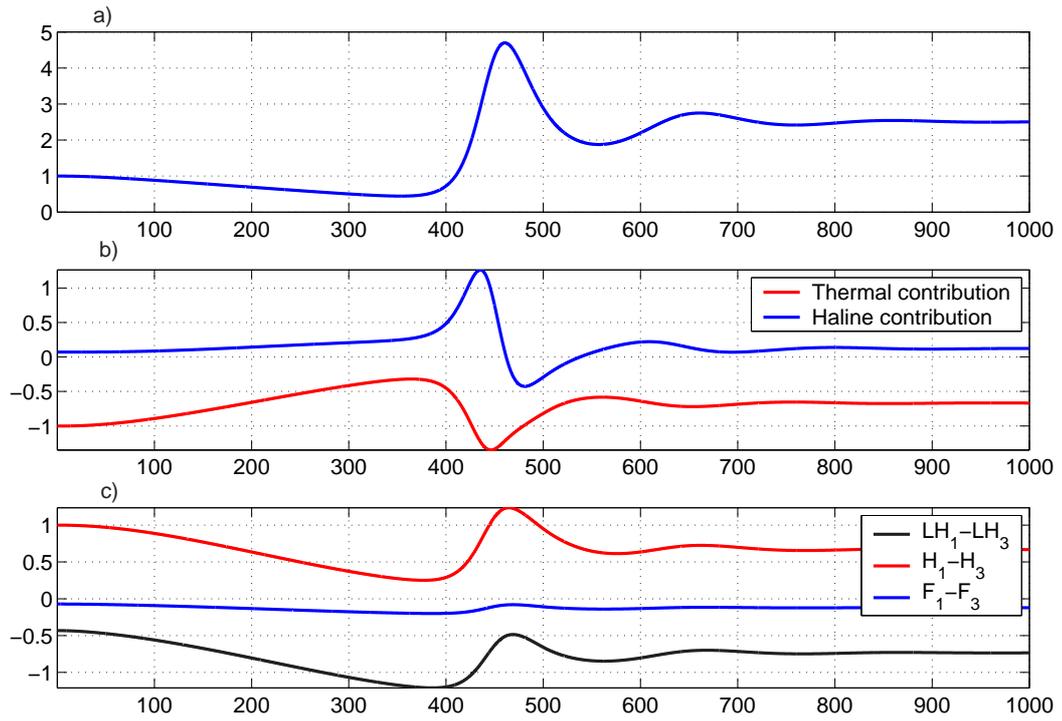

Figure 3: a) Evolution of the THC strength in units of $q(0)$ for the subcritical case presented in figure 1. b) Evolution of the thermal and haline oceanic advection forcings to the strength of the THC in units of $|\alpha q(0)(T_2(0) - 2T_1(0) + T_3(0))|$. c) Evolution of the most relevant atmospheric contributions to the forcing to the strength of the THC. Same units as in b). Abscissae: elapsed time expressed in $y$.

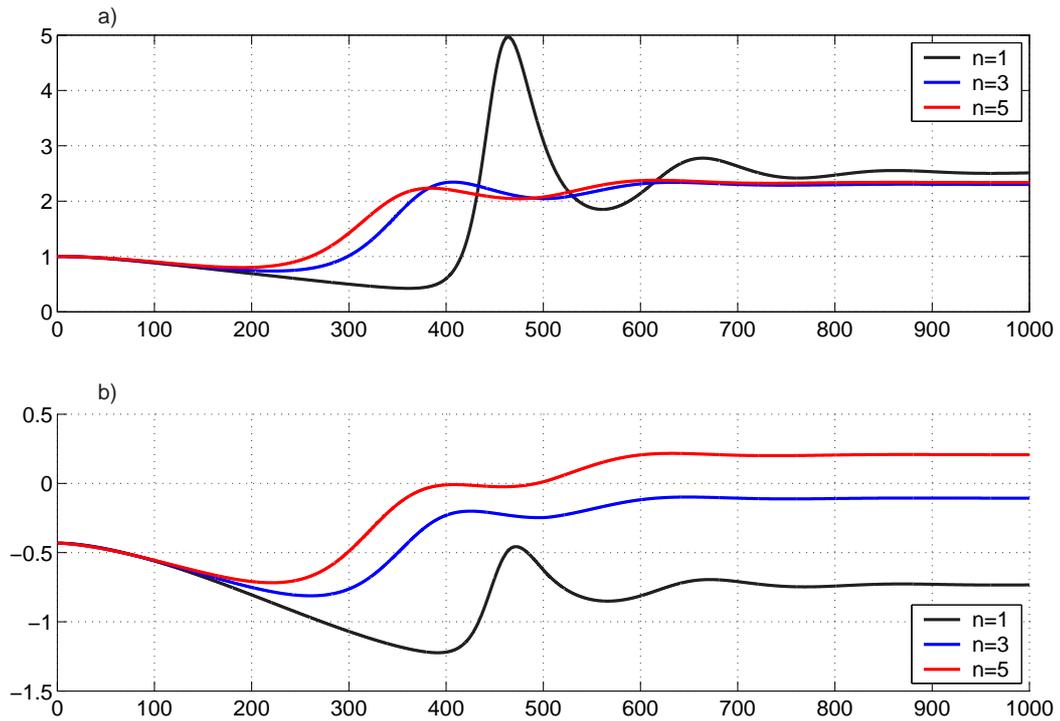

Figure 4: a) Evolution of the THC strength in units of $q(0)$ for various values of $n$. Radiative forcing is as in the subcritical case presented in figure 1. b) Evolution of the latent heat fluxes contributions $LH_1 - LH_3$ to the forcing for various values of $n$ in units of $|\alpha q(0)(T_2(0) - 2T_1(0) + T_3(0))|$. Abscissae: elapsed time expressed in $y$.

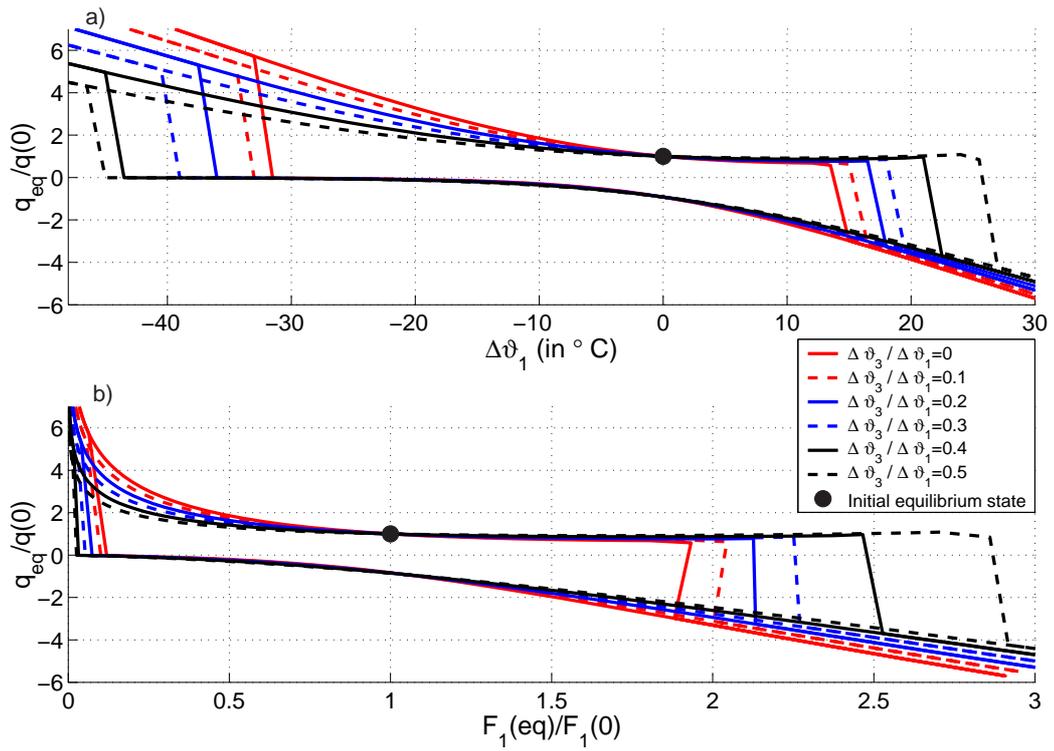

Figure 5: Hysteresis graphs of the system with $n = 1$ for radiative temperature perturbations. a) Radiative temperature change $\Delta\vartheta_1$ is considered as the relevant parameter. b) Freshwater flux $F_1$ is considered as the relevant parameter.

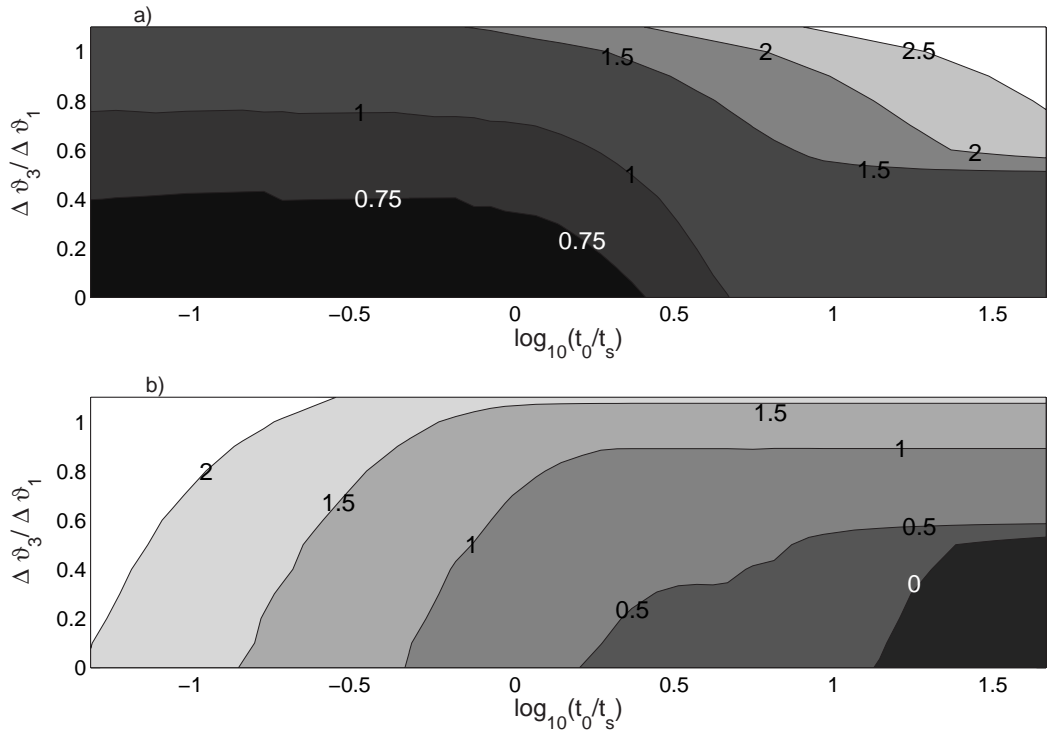

Figure 6: Stability of the THC against perturbations to the radiative temperature with $n = 1$. a) Critical values of the total increase of the freshwater flux; contours of $\log_{10}[\Delta\vartheta_1]$. $\Delta\vartheta_1$ is in $°C$. b) Critical values of the rate of increase of the freshwater flux; contours of $\log_{10}[\vartheta_1^t]$. $\vartheta_1^t$ is in $°C t_s^{-1}$, where $t_s$ is the advective time.

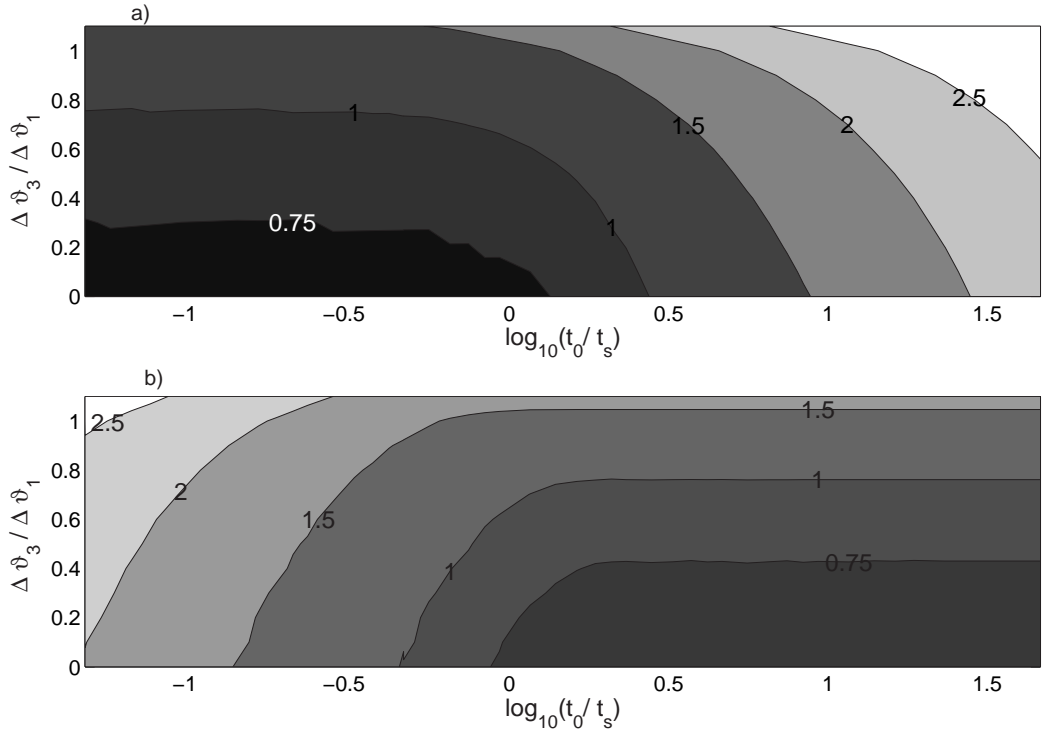

Figure 7: Stability of the THC against perturbations to the radiative temperature with $n = 3$. a) Critical values of the total increase of the freshwater flux; contours of $\log_{10}[\Delta\vartheta_1]$. $\Delta\vartheta_1$ is in $°C$. b) Critical values of the rate of increase of the freshwater flux; contours of $\log_{10}[\vartheta_1^t]$. $\vartheta_1^t$ is in $°C t_s^{-1}$, where $t_s$ is the advective time.

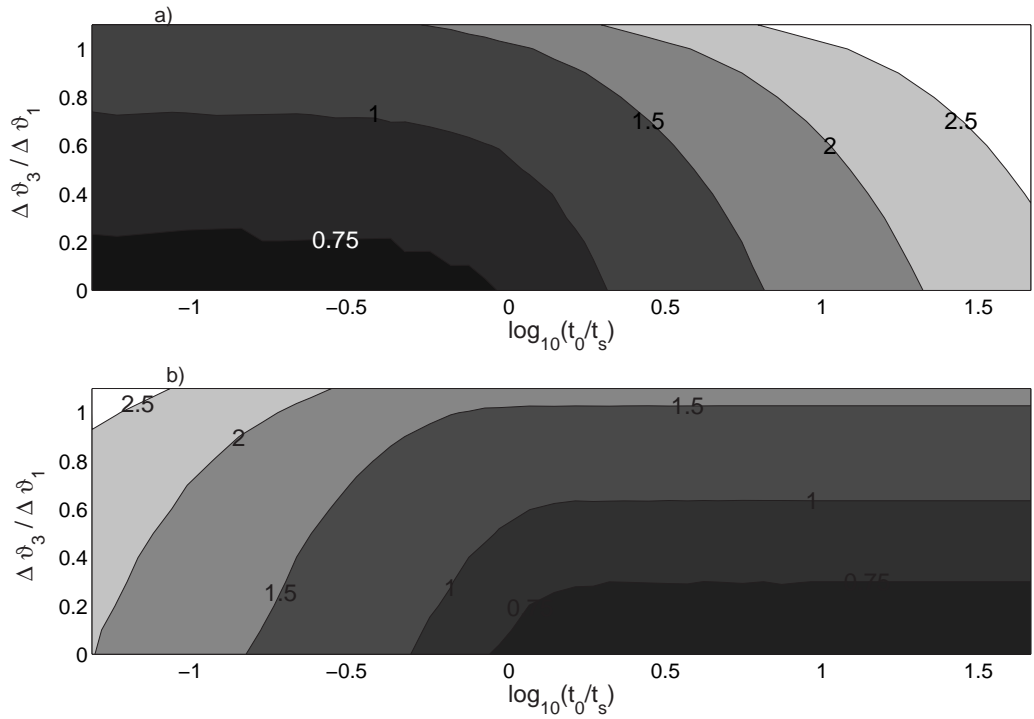

Figure 8: Stability of the THC against perturbations to the radiative temperature with $n = 5$. a) Critical values of the total increase of the freshwater flux; contours of $\log_{10}[\Delta\vartheta_1]$. $\Delta\vartheta_1$ is in $°C$. b) Critical values of the rate of increase of the freshwater flux; contours of $\log_{10}[\vartheta_1^t]$. $\vartheta_1^t$ is in $°C t_s^{-1}$, where $t_s$ is the advective time.

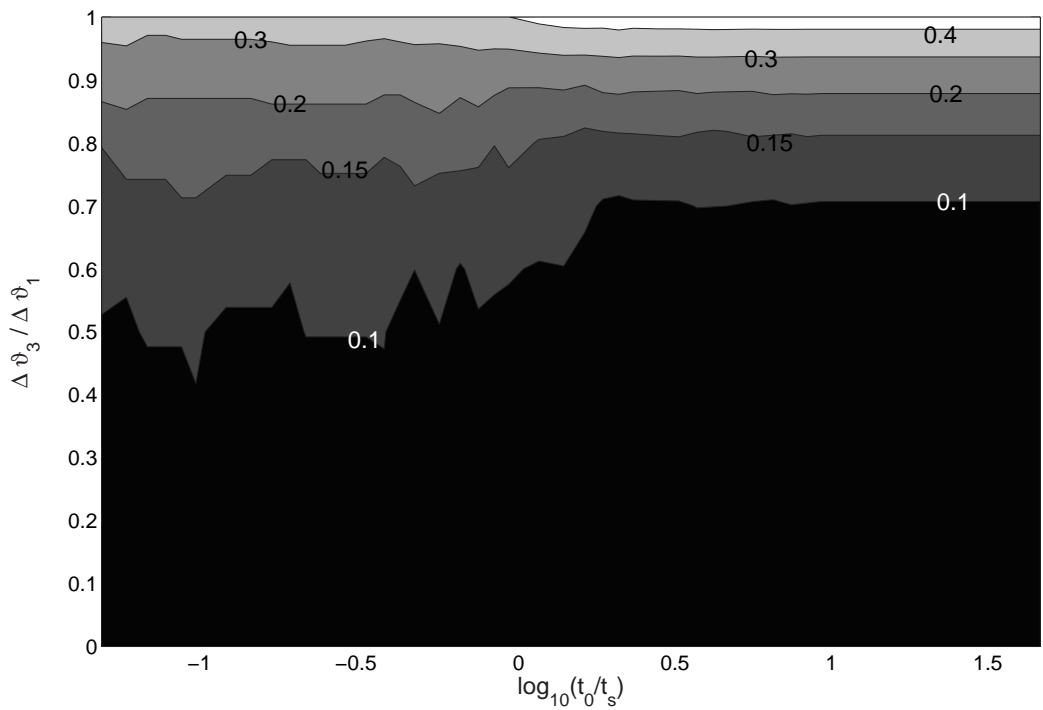

Figure 9: Sensitivity to the low-to-high latitudes radiative forcing ratio $TRRF$ of of the critical values of the total increase of the radiative temperatures; contours of $\log_{10}\left[Z_C\left(TRRF=1.0\right)/Z_C\left(TRRF=2.0\right)\right]$.

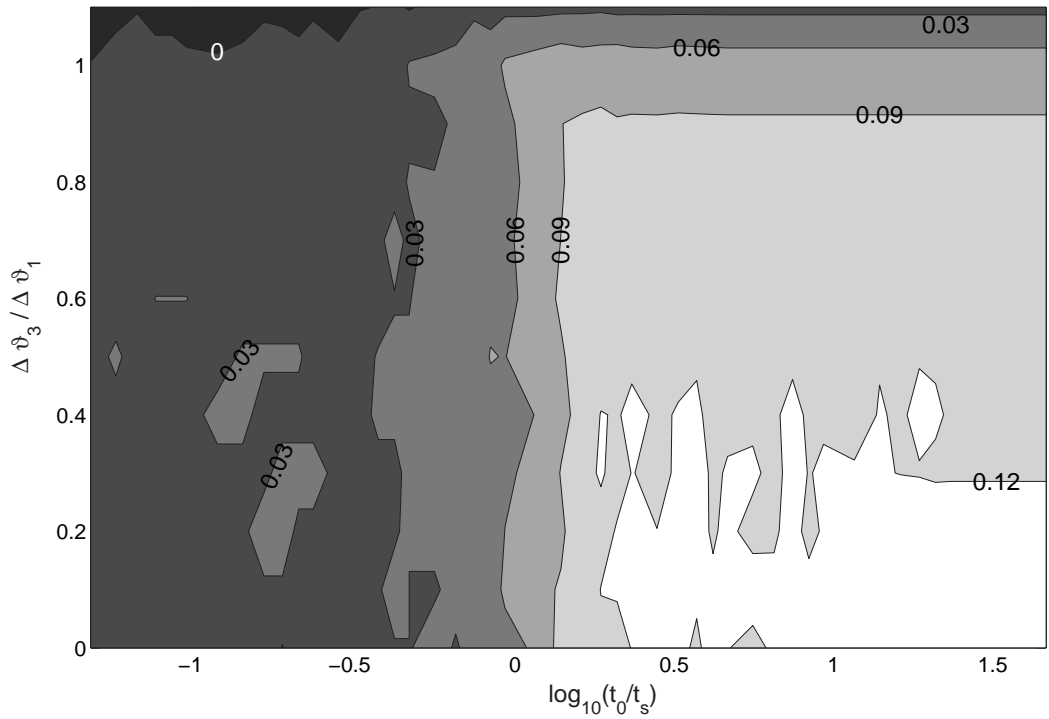

Figure 10: Sensitivity to the atmospheric transport parameterization of of the critical values of the total increase of the radiative temperatures; contours of $\log_{10}\left[Z_C\left(n=5\right)/Z_C\left(n=3\right)\right]$.